\documentclass[aps,pre,superscriptaddress,showpacs,showkeys,twocolumn,floatfix,nofootinbib,nobibnotes]{revtex4}

\usepackage{graphicx}

\begin{document}

%\pagestyle{headings}{Echeverr\'\i a \textit{et. al.}}
%\leftmark{Echeverr\'\i a \textit{et. al.}}

%\preprint{}

\title{Interpretation of heart rate variability via detrended fluctuation analysis and $\alpha\beta$ filter}

\author{J. C. Echeverr\'\i a}
\email[E-mail: ]{jcea@xanum.uam.mx}
\affiliation{School of Electrical and Electronic Engineering. University of Nottingham, Nottingham U.K.}
\affiliation{Electrical Engineering Department. Universidad Aut\'onoma Metropolitana-Izt., M\'exico City}
%\footnote{jcea@xanum.uam.mx}
\author{M. S. Woolfson}
\affiliation{School of Electrical and Electronic Engineering. University of Nottingham, Nottingham U.K.}
\author{J. A. Crowe}
\affiliation{School of Electrical and Electronic Engineering. University of Nottingham, Nottingham U.K.}
\author{B. R. Hayes-Gill}
\affiliation{School of Electrical and Electronic Engineering. University of Nottingham, Nottingham U.K.}
\author{G. D. H. Croaker}
\affiliation{Nottingham City Hospital, Nottingham U.K}
\affiliation{Queens Medical Centre, Nottingham U.K.}
\author{H. Vyas}
\affiliation{Queens Medical Centre, Nottingham U.K.}

\date{\today}

%%%%%%%%%%%%%%%%%%%%%%%%%%%%%%%%%%
\begin{abstract}
Detrended fluctuation analysis (DFA), suitable for the analysis of nonstationary time series, has confirmed the existence of persistent long-range correlations in healthy heart rate variability data. In this paper, we present the incorporation of the $\alpha\beta$ filter to DFA to determine patterns in the power-law behaviour that can be found in these correlations. Well-known simulated scenarios and real data involving normal and pathological circumstances were used to evaluate this process. The results presented here suggest the existence of evolving patterns, not always following a uniform power-law behaviour, that cannot be described by scaling exponents estimated using a linear procedure over two predefined ranges. Instead, the power law is observed to have a continuous variation with segment length. We also show that the study of these patterns, avoiding initial assumptions about the nature of the data, may confer advantages to DFA by revealing more clearly abnormal physiological conditions detected in congestive heart failure patients related to the existence of dominant characteristic scales.
\end{abstract}

\pacs{87.10.+e, 87.19.Hh, 87.19.Xx, 87.80.Tq}
\keywords{HRV, \textit{1/f}, DFA,  $\alpha\beta$ filter, congestive heart failure}

\maketitle

%\newpage
\section{Lead Paragraph}

Detrended fluctuation analysis (DFA) \cite{4}, suitable for the analysis of nonstationary time series, has confirmed the existence of persistent long-range correlations in healthy heart rate variability (HRV) data \cite{1}. Notwithstanding the early success in the application of DFA, a limitation exists concerning the final stage of the technique as applied to HRV. This is related to the calculation of the power-law scaling exponent for the short-term or long-term range, a range division that has been assumed by the presence of the ``crossover phenomenon'', or critical segment length where there is a sudden change in the power law, found in several patients \cite{4,10}. We believe that one should not reduce DFA to quantify only two (a short- and a long-range) scaling exponents, but instead determine if there is more information in the power-law behaviour. A previous attempt using coarse local slopes for the scaling exponent has already suggested a kind of scaling stability impairment with heart disease \cite{11}; however, a detailed analysis of the patterns of the slope is not provided. Here, we present the incorporation of the $\alpha\beta$ filter to the DFA for the estimation of the power law as a function of the time scales. The $\alpha\beta$ filter \cite{Kingsley} is a recursive least-squares method that has been used for tracking targets or to characterise the operation of an induction motor drive, for example \cite{12,13}. We have employed this filter as a more appropriate approach to obtain a detailed analysis of the scaling behaviour by improving the precision in the interpretation of the DFA results. Well-known simulated scenarios and real HRV data involving normal and pathological circumstances were used to evaluate this enhancement. Our results suggest the existence of evolving scaling patterns not always presenting a uniform power-law behaviour that can be readily detected by a linear slope estimation procedure over two predefined ranges. We also show that the incorporation of the study of these patterns may confer advantages to the DFA by revealing more clearly abnormal physiological conditions detected in congestive heart failure (CHF) patients related to the existence of dominant characteristic scales.

\section{Introduction}

In recent years, power spectral techniques have been widely applied to short-term fluctuations of the elapsed time between consecutive heart beats or RR intervals. Although disagreement exists about the physiological information provided by the frequency bands \cite{amjphc02,karama,eckberg}, it is accepted in some circles that these tools have revealed unique spectral signatures corresponding to the autonomic control dynamics \cite{1,amjphc02}. Recently, this spectral quantification of the magnitude of this HRV has been complemented with long-memory statistical measurements \cite{2} revealing, for healthy dynamics, persistent long-range correlations over a wide range of time scales \cite{3}. This has promoted the adoption of a physiological conjecture suggesting that HRV, lacking characteristic scales and having long-range correlation, indicates a healthy organising principle that seems to break down in several pathological states. For example, a disruption of long-range variability correlations, perhaps reflecting a shift to a dominant single characteristic time scale, has been detected in the heartbeat dynamics of CHF patients \cite{3,4}. 

For the analysis of long-term HRV that lacks characteristic scales it is appropriate to apply a fractal approach to identify the existence of persistent correlations over a wide range of time scales. Moreover, the introduction of DFA \cite{4,5,IvaChaos} has conferred advantages over other conventional fractal methods because it permits the detection of long-range correlations embedded in nonstationary time series and avoids spurious detection of apparent long-range correlations that are artifacts of nonstationarity behaviour \cite{4,6,7,8}. Any invariant scaling characteristics in the heart rate fluctuations obtained by these means can therefore be mostly attributed to the intrinsic mechanism of neuroautonomic control \cite{8} with the advantage of not having to rigorously control physical activity or provide an external stimulus, hence promoting the possibility of ambulatory monitoring for long-term evaluation. The DFA has also been applied to detect long-range correlations in other time series like heterogeneous DNA sequences and \cite{5,9} the stride interval fluctuations obtained from unconstrained human gait dynamics \cite{3}.

\section{DFA and $\alpha\beta$ filter}

The DFA method has been described elsewhere \cite{4}. Briefly, the original RR 
interval or HRV series is initially integrated as follows:

\begin{equation}
\label{eqnew}
Y (k) =  \sum_{i=1}^k [RR(i)-RR_{ave}]
\end{equation}

\noindent
where $RR(i)$ is the \textit{i}th RR interval and $RR_{ave}$ is the average RR interval of the series. This integration is used to overcome both the restricted observation time and physical nature of the process. Next, the integrated series is divided into windows, or boxes, of equal numbers of $n$ RR intervals. In each window, the local trend is obtained by least-squared line 
fit\footnote{Higher order polynomials may also be used in this fitting procedure. However, it has been reported that the inherent deviations at small scales appear to become stronger in higher order of DFA \cite{14,Hu}. Additionally, crossovers produced by sinusoidal trends, like the ones analysed in this contribution, are shifted to larger values for higher orders of detrending \cite{14,Hu}. Finally, HRV data have been normally analysed using linear detrending DFA which is the order that we employed in this study.}. This trend is locally subtracted from the integrated series to reduce the nonstationary 
artifacts. The average root-mean-square fluctuation, $F_{m}(n)$, is then 
calculated. The previous procedure is repeated for all window sizes (time scales). 
Consequently, the relationship on a double-log graph between these 
fluctuations $F_{m}(n)$ and time scales $n$ can be approximately evaluated by a 
linear model: $F_{m}(n)\sim n^{\zeta}$ . A resulting slope, or scaling exponent 
$\zeta$, of 0.5 indicates white noise and the absence of long-range 
correlations, an exponent of 1 reflects the behaviour of a $1/f$ process having 
persistent long-range correlations whilst the slope of 1.5 indicates a 
random walk (Brownian noise) with a very smooth behaviour where correlations still exist but not in a power-law form\footnote{In spite of 
recognising that $\alpha$ has been the notation extensively used elsewhere to describe the fractal scaling exponent, we have used instead the notation $\zeta$ to avoid creating confusion with the $\alpha\beta$ filter concept described in this paper.}.

The scaling exponent of the average root-mean-squared fluctuations as a 
function of time scales is usually estimated by the slope of the double-log 
plot covering short- or long-term ranges. Rather than finding the scaling 
exponent only for these two generally predefined ranges, in this paper we 
have adopted the $\alpha\beta$ filter \cite{12, 13, Kingsley, Mal_GEC} to recursively estimate a local least-squares fitting for tracking the evolution of the gradient (\textit{i.e.}, of the power law) as a function of log time scales (or window sizes). This gradient is also referred to in this paper as \textit{scaling pattern}. 

Essentially, the $\alpha\beta$ filter \cite{Kingsley} is a simplified version of a Kalman filter that provides a good compromise between performance and computational load \cite{Mal_GEC}. By adopting a recursive approach to the estimate of DFA scaling patterns we have assumed a constant local slope in these patterns. Obviously, it is necessary to consider that this slope may involve variations from linearity. These variations are modelled in the Kalman filter as plant noise \cite{Mal_GEC}. By contrast, in the $\alpha\beta$ filter, plant noise is implicitly assumed to be zero \cite{Kingsley,Mal_GEC}. However, the effects of plant noise may also be simulated by preventing the $\alpha\beta$ filter gains from going to zero (see equations \ref{eq6}--\ref{eq7})\cite{Mal_GEC}. Thus, it is not necessary to know the exact model of the local slopes and consequently the $\alpha\beta$ filter can be used with the advantage of reducing the number of required operations. A brief description of how the $\alpha\beta$ filter was incorporated to DFA (equations \ref{eq1}--\ref{eq7}) is presented next. 

Let $G_{m}(n)$ be the log of the average root-mean-squared fluctuations, 
$F_{m}(n)$, produced by DFA at the window size $n$, with $n$ being the number of RR 
intervals (or the equivalent physiological time scale). Define 
$m_{e}(n)$ as the required estimate of the gradient at the log window size $n$. 
Since a logarithmic representation of the window size $n$ is employed, an 
initial interpolation of $G_{m}(n)$ is convenient to have it parameterised by a 
uniformly sampled variable in the logarithmic representation of the window 
size domain; using $k$ as an index to denote the discrete elements or samples of this new variable. Then, it is possible to predict the value of the 
root-mean-squared fluctuations at $k$ according to

\begin{equation}
\label{eq1}
G_p (k) = G_e (k - 1) + m_e (k - 1)\delta 
\end{equation}

\noindent
where $G_{p}(k)$ is the predicted log root-mean-squared fluctuations, 
$G_{e}(k-1)$ is the estimate of log root-mean-squared fluctuations at $k-1$, and 
$\delta$ is the uniform separation between successive elements of the parameterised 
variable. The original piece of information, \textit{i.e.}, the log root-mean-squared 
fluctuations $G_{m}(k)$, produced by the DFA method at $k$, is combined with the 
previous prediction $G_{p}(k)$ according to

\begin{equation}
\label{eq2}
G_e (k) = [1 - \alpha (k)]G_p (k) + \alpha (k)G_m (k)
\end{equation}

\noindent
in order to obtain the estimate of the log root-mean-squared fluctuations at 
$k$. Here, $\alpha (k)$ is a smoothing coefficient at $k$. Finally, the desired estimate of 
the gradient at $k$, $m_{e}(k)$, can be obtained from

\begin{equation}
\label{eq3}
m_e (k) = m_e (k - 1) + \frac{\beta (k)}{\delta }[G_m (k) - G_p (k)]
\end{equation}

\noindent
where $\beta (k)$ is another smoothing coefficient. The smoothing coefficients $\alpha (k)$ and $\beta (k)$ are given by

\begin{equation}
\label{eq4}
\alpha (k) = \frac{2(2k - 1)}{k(k + 1)}
\end{equation}

\noindent
and

\begin{equation}
\label{eq5}
\beta (k) = \frac{6}{k(k + 1)}
\end{equation}

To take into account any possible deviation from linearity in the log domain 
(perhaps owing to oscillatory trends \cite{14} or to finite-length effects \cite{15}) 
in the root-mean-squared fluctuations produced by the DFA method, the 
coefficients $\alpha (k)$ and $\beta (k)$ are prevented from going to zero for $k > Q$ as follows:

\begin{eqnarray}
\alpha (k) = \alpha (Q) \label{eq6}\\
\beta (k) = \beta (Q)	\label{eq7}
\end{eqnarray}

\noindent 
where for our analysis we empirically found a value $Q$=500 as a balance between 
``over smoothing'' at high $Q$ where changes in the gradient are not adequately 
tracked, and ``under smoothing'' where the effects of noise become 
significant.

\section{Data to be analysed}

The incorporation of this $\alpha\beta$ filter to the DFA was initially explored with simulated data having 100800 samples (approximately corresponding to the 
amount of RR intervals recorded during 24 hours at 70 beats per minute). 
These simulations involved Gaussian white noise series; Brownian motion 
series (obtained by the integration of the white noise series \cite{16}); $1/f^{\gamma}$ 
process ($\gamma=1$) generated by the spectral synthesis approximation suggested by Suape \cite{16}; and finally $1/f$ series including superimposed oscillatory trends introduced by the addition to the previous $1/f$ series of a pure sine component with 0.01 Hz frequency and an amplitude of 0.1.

In addition, the above mentioned methods were applied to 18 normal sinus HRV long-term records (20-24 hours) from subjects without diagnosed 
cardiac abnormalities as well as 15 long-term records (20 hours) from 
subjects with severe CHF. The unaudited RR annotation 
marks of these records were gathered from PhysioNet \cite{17}. The CHF records 
have also been used elsewhere \cite{4} to explore the DFA capabilities. This 
database includes long-term ECG recordings from 15 subjects (11 men, aged 22 
to 71, and 4 women, aged 54 to 63). The normal sinus heart rate database was 
obtained from subjects found to have no significant arrhythmias; they 
included 5 men, aged 26 to 45, and 13 women aged 20 to 50 \cite{17}. The 
annotation marks were analysed to remove unqualified sinus beats, these 
outliers being detected according to the method suggested in Ref. \cite{18}, which 
relies upon confirming each RR interval to be within a tolerance range 
covering the 20{\%} extremes from a running mean value obtained through the 
4 adjacent intervals. Only recordings with more than 85{\%} qualified beats 
were included in the analysis as suggested in Ref. \cite{19}. Thus, records CHF02 and 
CHF06 were excluded from the analysis of the CHF database. 

Finally, since it has been reported that day-night differences exist in the scaling behaviour of HRV data \cite{8}, the enhanced DFA analysis (\textit{i.e.}, DFA plus $\alpha\beta$ filter) was also applied separately to wake and sleep 6 hours segments of normal and CHF data.

\section{Results and Discussion}

\begin{figure*}
\includegraphics{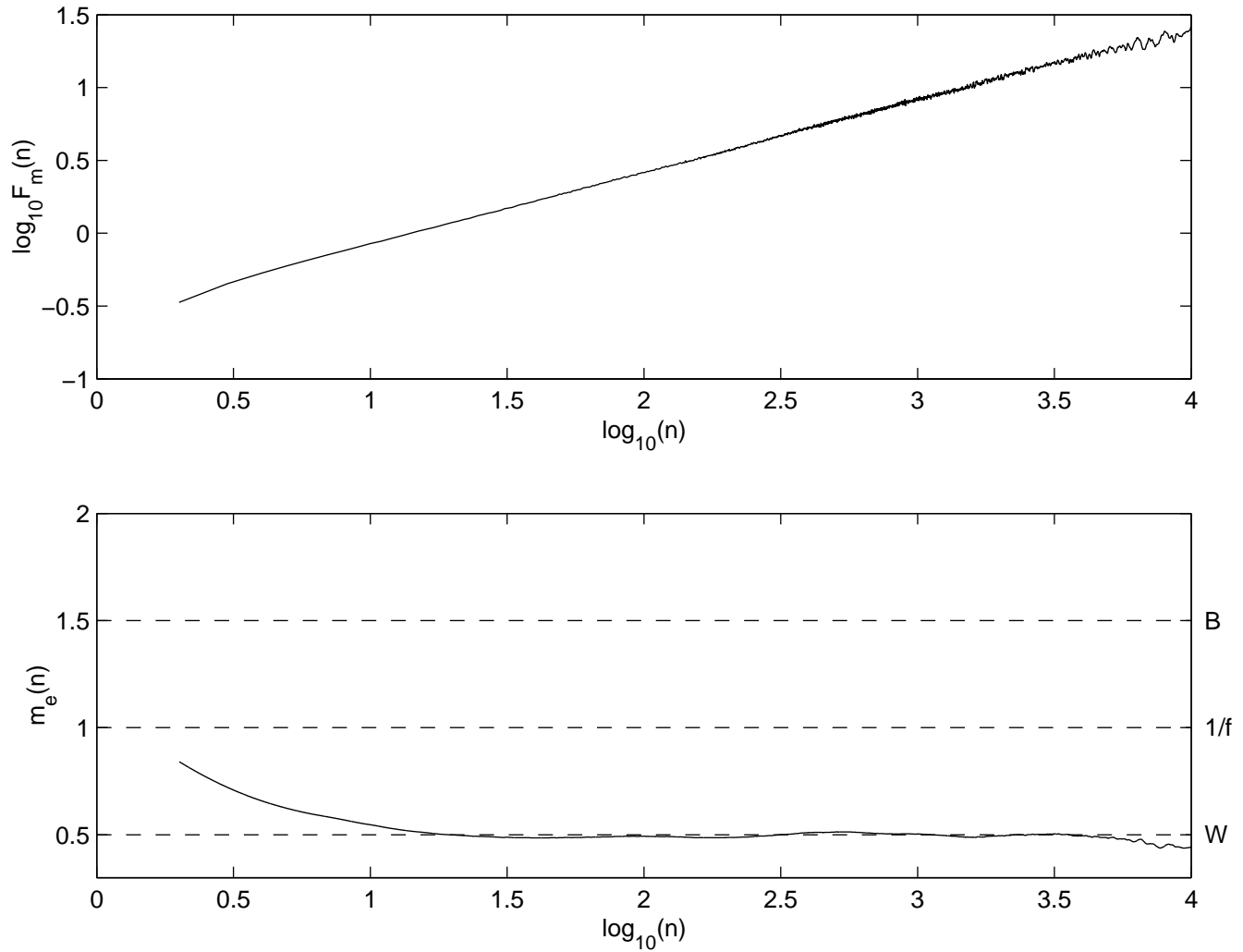}
\caption{\label{fig1a}Result of DFA and $\alpha\beta$ filter applied to white noise series. Top plot shows log $F_{m}(n)$ \textit{vs} log \textit{n} obtained via DFA whilst the bottom plot presents $m_{e}(n)$ \textit{vs} log $n$ 
estimated by $\alpha\beta$ filter. Values of log $n$ go up to 4 to avoid presenting larger deviations resulting from the finite-length effects. Horizontal dashed lines in the bottom plot indicate the scaling exponent values 0.5 (W), 1 (1/f), and 1.5 (B) 
corresponding to white noise, $1/f$ process, and Brownian noise, respectively.}
\end{figure*}

\begin{figure*}
\includegraphics{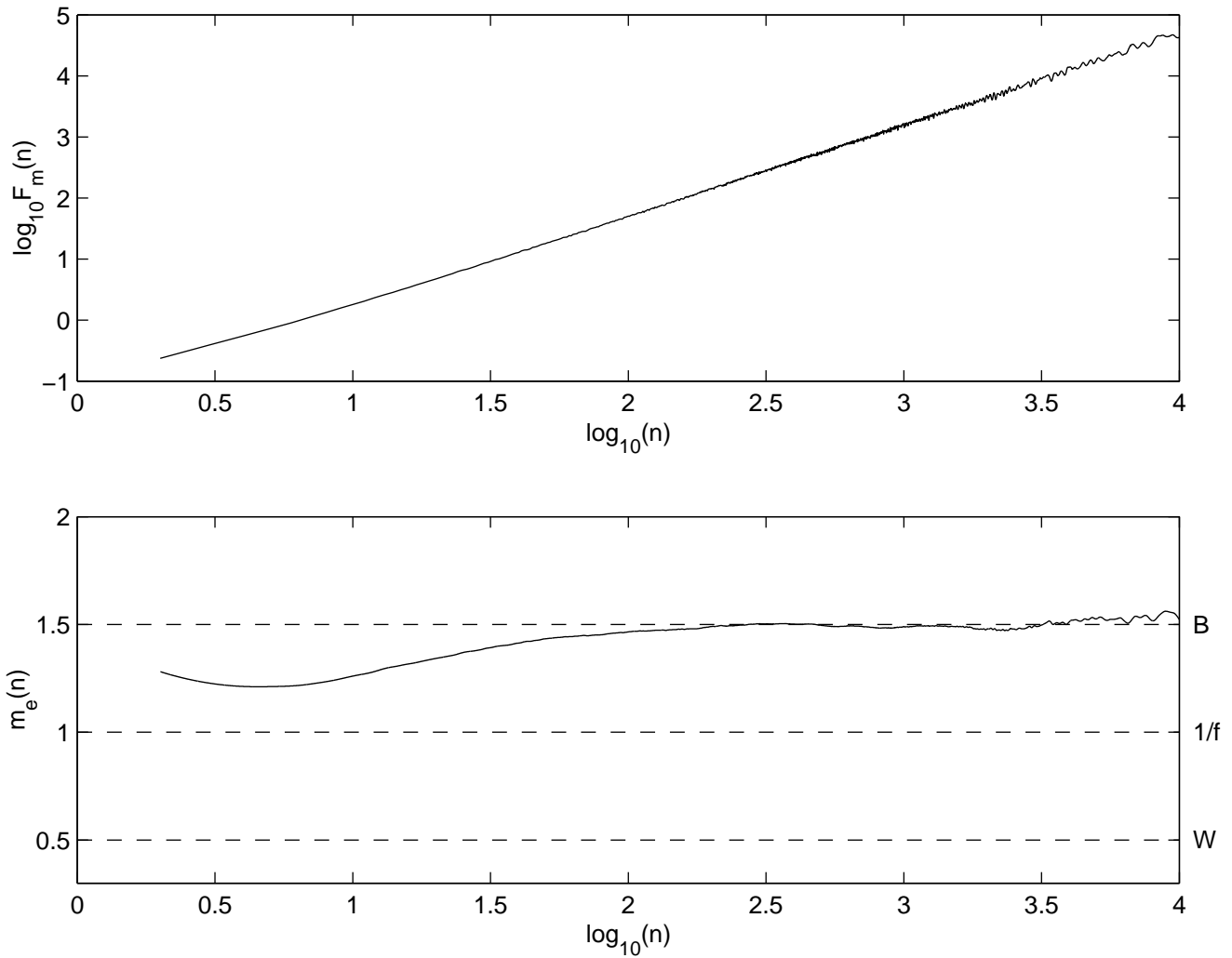}
\caption{\label{fig1b}Result of DFA and $\alpha\beta$ filter applied to Brownian motion. The top plot shows log $F_{m}(n)$ \textit{vs} log \textit{n} obtained via DFA whilst the bottom plot presents $m_{e}(n)$ \textit{vs} log $n$ 
estimated by $\alpha\beta$ filter.}
\end{figure*}

\begin{figure*}
\includegraphics{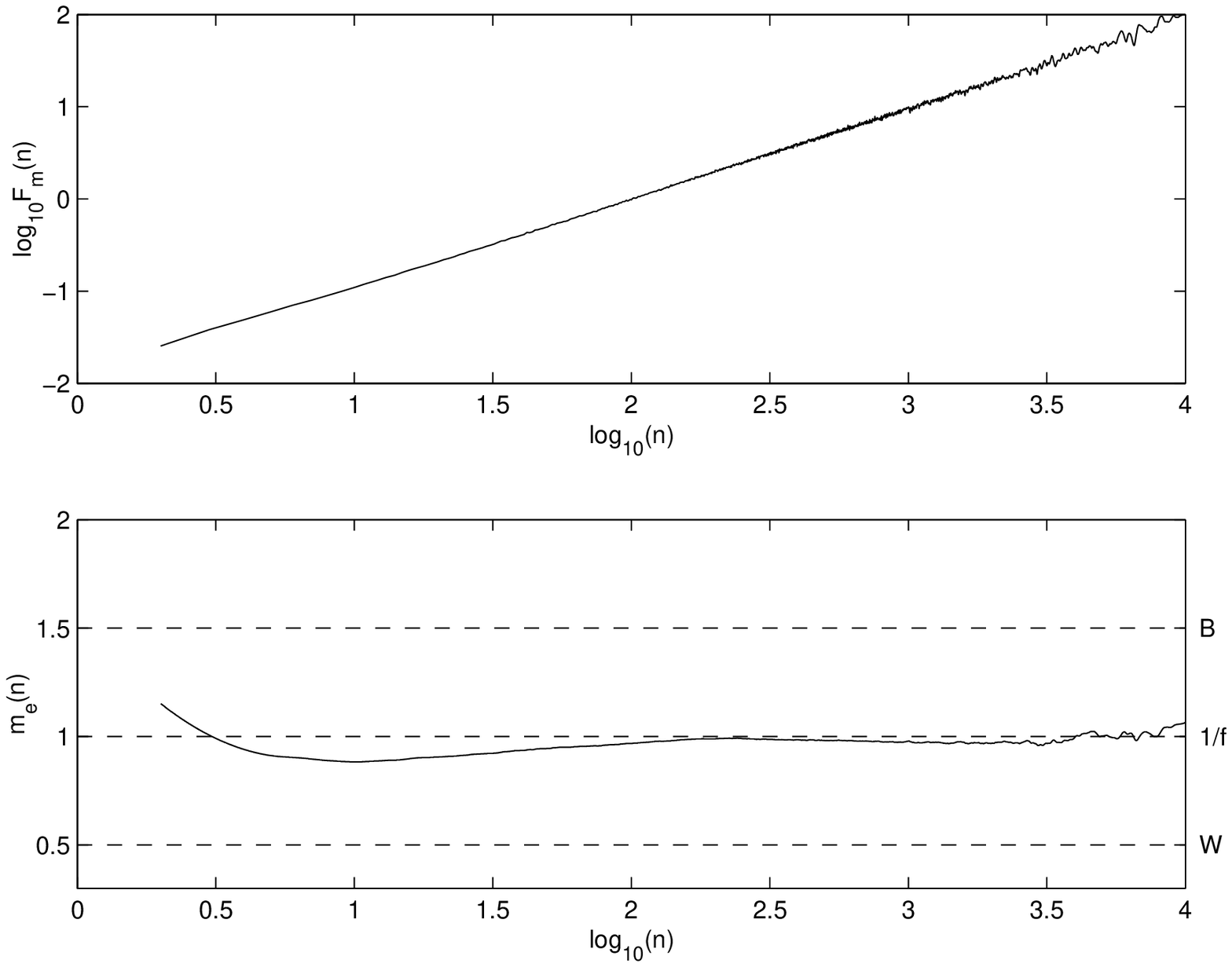}
\caption{\label{fig1c}Result of DFA and $\alpha\beta$ filter applied to $1/f$ process. The top plot shows log $F_{m}(n)$ \textit{vs} log \textit{n} obtained via DFA whilst the bottom plot presents $m_{e}(n)$ \textit{vs} log $n$ estimated 
by $\alpha\beta$ filter.}
\end{figure*}

Figures \ref{fig1a}, \ref{fig1b}, and \ref{fig1c} present the result of the DFA and $\alpha\beta$ filter applied to 
the following simulated series: Gaussian white noise, Brownian motion, and 
$1/f$ noise. The top of each figure shows the double-log plot of the average 
root-mean fluctuations $F_{m}(n)$ as a function of the window size $n$ obtained 
using DFA whilst the bottom figure corresponds to the gradient 
$m_{e}(n)$ of these fluctuations, again as a function of the log window size, 
estimated by the $\alpha\beta$ filter. Note that although the original series length was 100800 samples, corresponding to the log range of 5, these plots cover window sizes up to log value 4 to avoid presenting the larger deviations resulting from the finite-length effects of the data \cite{10, 15} (which can be also explained as being related to averaging over too few windows giving only rough detrending with the effects of noise becoming more significant). These type of results have been reported elsewhere \cite{Hu}, but for the purpose of this contribution they are useful to promote familiarisation with the type of scaling patterns obtained via the $\alpha\beta$ filter in well-characterised scenarios. As was expected for these simulations the scaling pattern reveals the theoretical value of 0.5 for 
white noise, 1.5 for Brownian noise, and 1 for the $1/f$ series. The gradient 
patterns in these figures confirm that the expected scaling behaviours were 
only approached asymptotically, as has been explained in Ref. \cite{14} because of the intrinsic deviations of the DFA method for small window sizes \footnote{It is important to mention that these deviations are not only intrinsic to DFA. For example, deviations from an expected scaling behaviour at small scales have also been described in the context of Hurst analysis \cite{Mandel}}. Additionally, as reported by Hu \textit{et. al.} \cite{Hu}, it is possible to appreciate that these deviations present larger scaling values than the expected behaviour for white noise whereas for the Brownian noise and $1/f$ series these deviations resulted in smaller values.

\begin{figure*}
\includegraphics{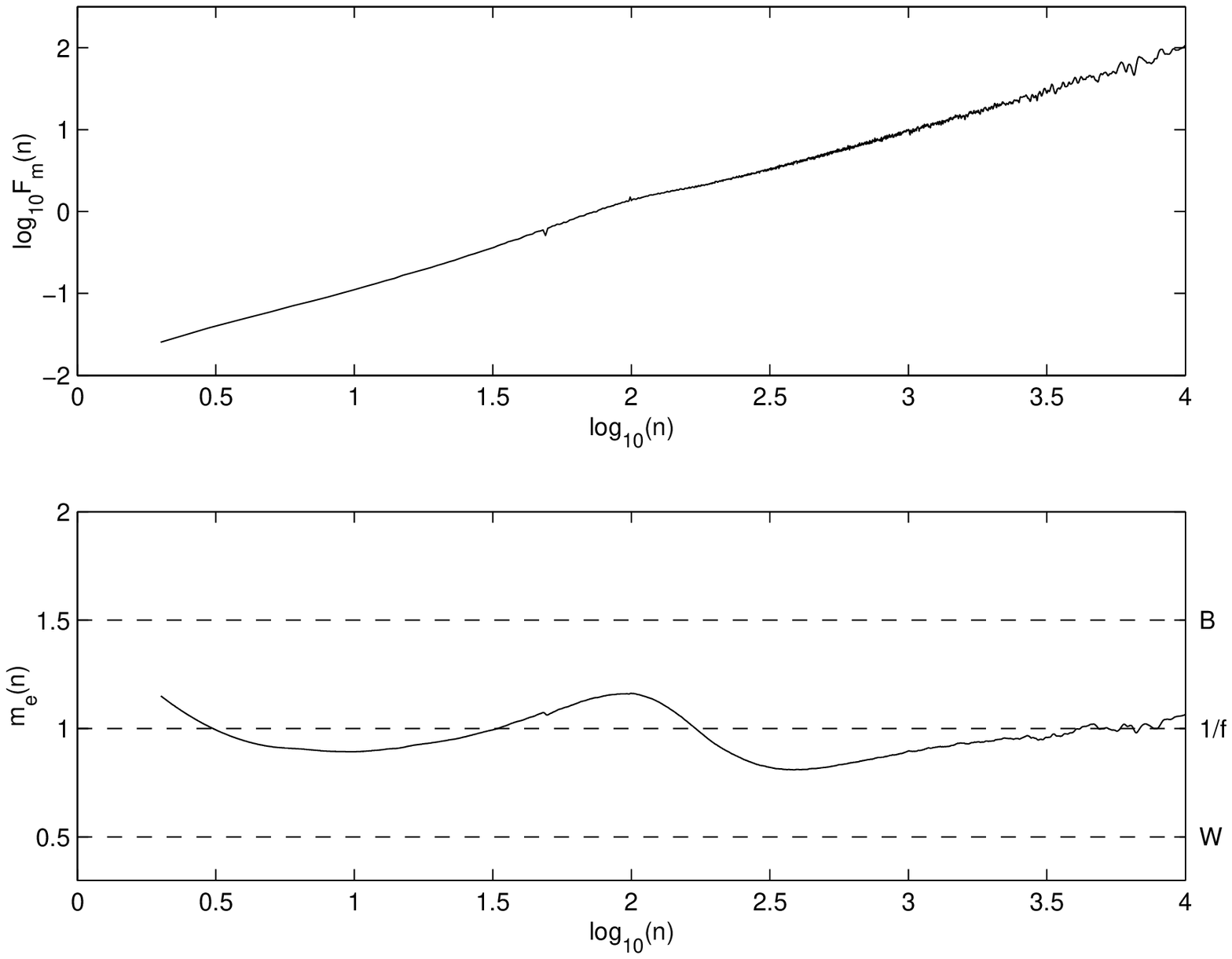}
\caption{\label{fig2}Result of DFA and $\alpha\beta$ filter applied to $1/f$ process corrupted by a sine component having 0.01 Hz of frequency and amplitude of 0.1. The top plot shows log $F_{m}(n)$ \textit{vs} log \textit{n} obtained via DFA whilst the bottom plot presents $m_{e}(n)$ \textit{vs} log $n$ estimated 
by $\alpha\beta$ filter.}
\end{figure*}

Figure \ref{fig2} presents the results of the DFA and $\alpha\beta$ filter applied to $1/f$ noise with a superimposed oscillatory trend. The estimated scaling pattern presented at the bottom of the figure clearly reveals a strong deviation in the linearity of the double-log plot (see Figure \ref{fig1c} for a comparison of the scaling behaviour 
obtained when analysing an uncorrupted $1/f$ process). The addition of 
this oscillatory trend has produced the type of ``crossover phenomenon'' 
suggested in Ref. \cite{4}. A reliable detection of this phenomenon is possible with 
the gradient, estimated by the $\alpha\beta$ filter, indicating a clear deviation 
from a uniform power law mostly because of a well-defined projection above unity within the log window size range of 1.5 to 2.2. This range includes the log window size 2 as the location presenting the 
major deviating effect. Hence, this is related to the superimposed trend having an oscillatory frequency of 0.01 Hz and a regular sampling frequency of 1 Hz (see also Kantelhardt \textit{et. al.} \cite{14} and Hu \textit{et. al.} \cite{Hu} for simulations and detailed analytical calculations of how a sinusoidal trend affects the scaling behaviour of correlated noise).

\begin{figure*}
\includegraphics{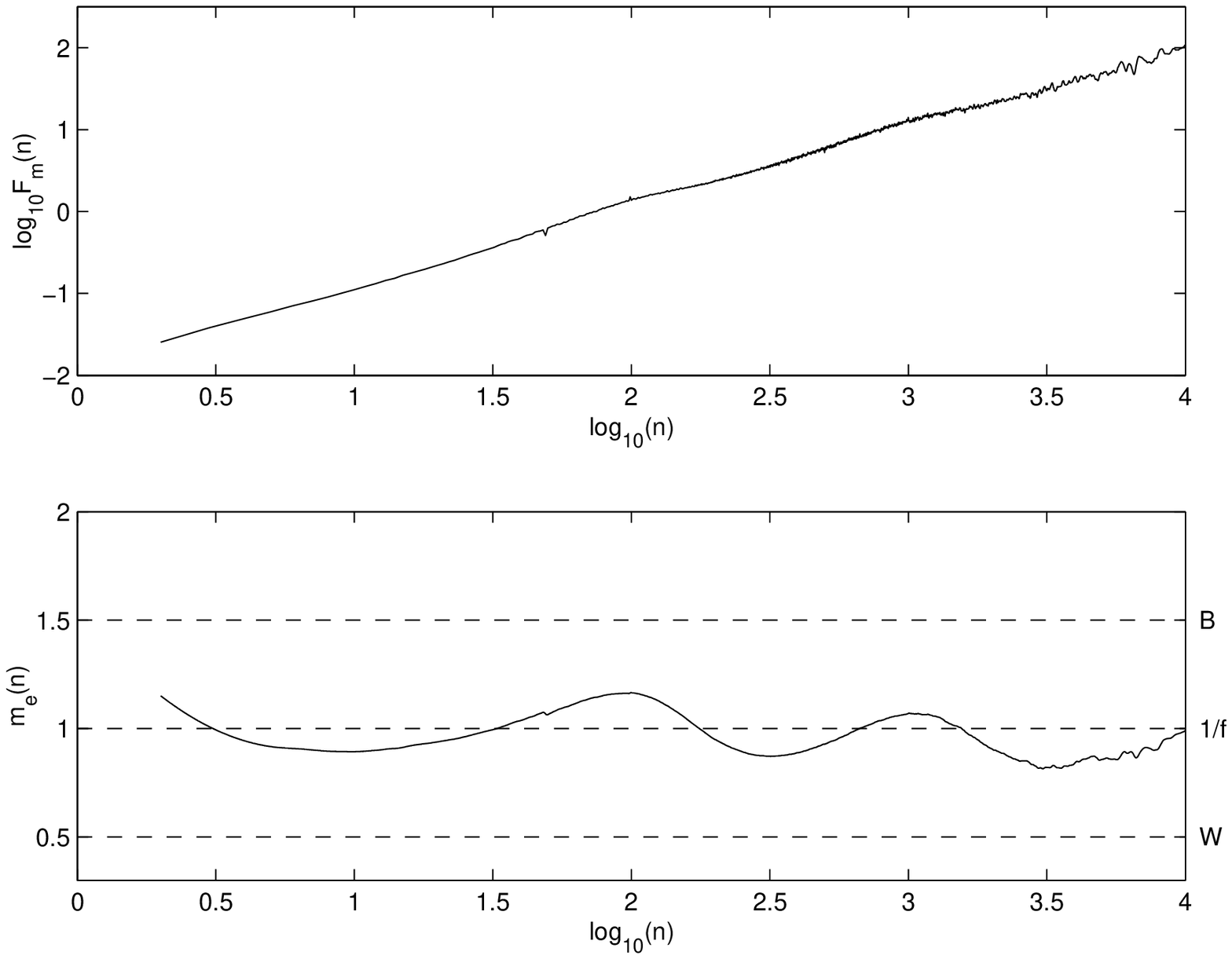}
\caption{\label{fig3}Result of DFA and $\alpha\beta$ filter applied to $1/f$ process corrupted by two sine components having 0.01 and 0.001 Hz of frequency and both having amplitude of 0.1. The top plot shows log $F_{m}(n)$ \textit{vs} log \textit{n} obtained via DFA whilst the bottom plot presents $m_{e}(n)$ \textit{vs} log $n$ estimated by $\alpha\beta$ filter.}
\end{figure*}

As suggested by the result of Figure \ref{fig2}, a complementary exploration of the 
addition of a second sine component of lower frequency (0.001 Hz) and same amplitude should reveal an additional projection in the log 3 
long-term range. The corresponding results for this scenario are presented in Figure \ref{fig3} where it is possible to confirm 
this prediction.

\begin{figure*}
\includegraphics{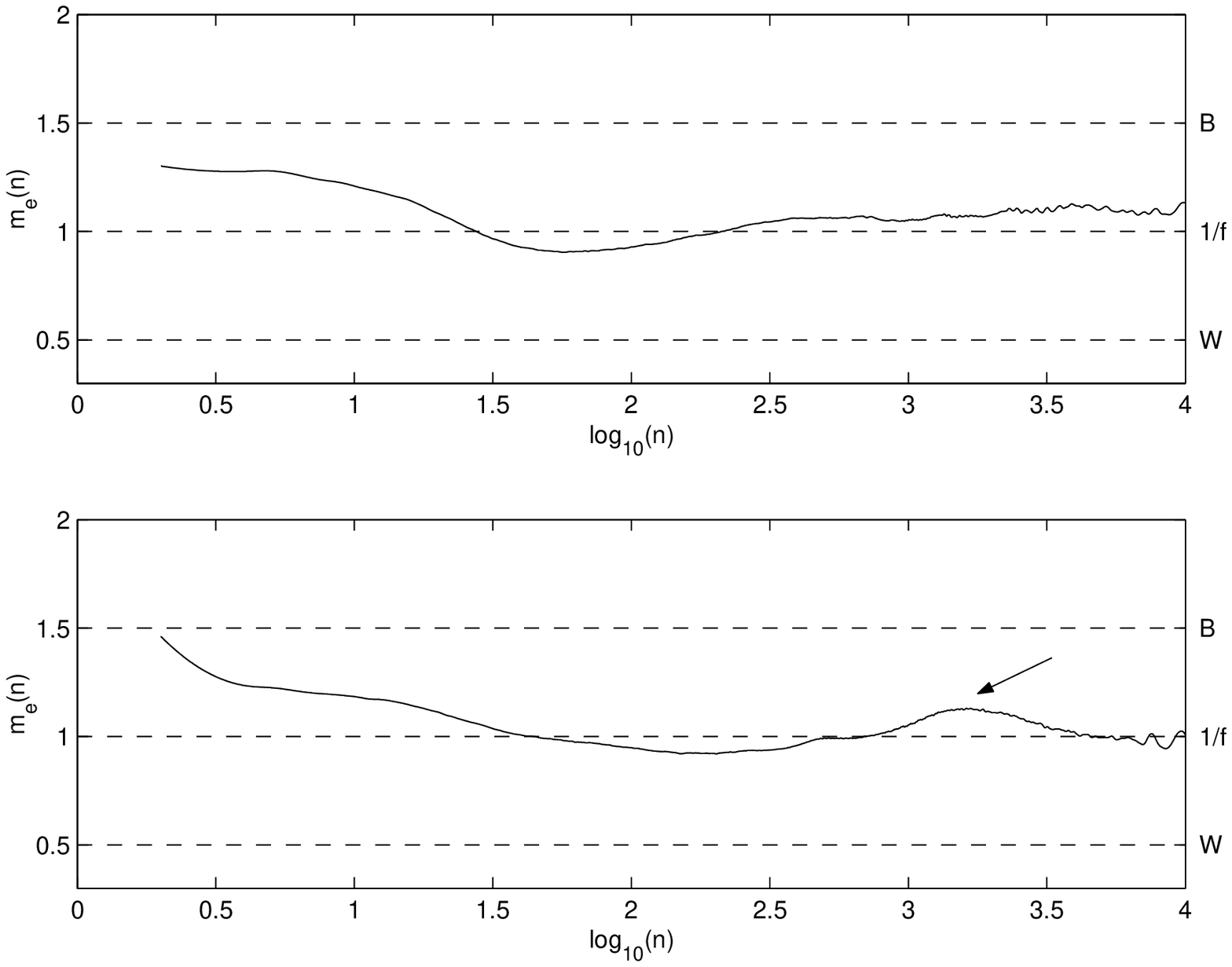}
\caption{\label{fig4_5}Result of DFA and $\alpha \beta$ filter applied to the RR interval series of Physionet normal patient 18184 and 19090. The top plot shows $m_{e}(n)$ \textit{vs} log $n$ estimated by the $\alpha\beta$ filter for patient 18184 whilst the bottom plot presents the same result for patient 19090. Note in this case the small projection in $m_{e}(n)$ indicated by an arrow in the long-term range.}
\end{figure*}

The top of Figure \ref{fig4_5}  presents a typical scaling pattern for a normal patient (18184). It 
is possible to describe two major features in the evolution of this pattern. For the long-term range, the power scaling pattern asymptotically 
attains a plateau having a value near unity; by analogy with the pattern 
obtained for the $1/f$ simulation (Figure \ref{fig1c}) it is clear that this result 
(top of Figure \ref{fig4_5}) suggests persistent correlations and the lack of characteristic scales in 
the long range in accordance with the hypothesis of healthy heart rate 
presenting a type of $1/f$ fluctuations \cite{3, 4}. In contrast, for the short time 
scales it is possible to appreciate deviations above unity, from a uniform 
power-law pattern, again in accordance with the results of Peng \textit{et. al.}, who have 
explained that this deviation is probably the result of the relatively smooth 
heartbeat oscillations, associated with respiration, that dominate the 
short-term range \cite{4}. 

For the case presented at the top of Figure \ref{fig4_5} it is clear, as reported in Ref. \cite{4}, 
that in principle it would be possible to arrive at the same kind of 
conclusions by assuming the existence of a crossover phenomenon. The use of the conventional linear quantification in predefined ranges at the top of Figure \ref{fig4_5} would assign the exponents for short-term ($n$ from 4 to 16) and long-term ($n$ from 100 to 10000) ranges of 1.09 and 1.10, respectively. However, the evolving scaling pattern produced by the $\alpha\beta$ filter, showing a considerable variation in the short-term range, indicates that an exponent of 1.09 is a gross approximation for this behaviour. Hence, further information may be revealed with this scaling pattern analysis that is not possible to detect by the conventional interpretation of the DFA results. Other advantages obtained by the incorporation of $\alpha\beta$ filter will be illustrated with the following results.

The bottom of Figure \ref{fig4_5} presents the scaling pattern for another normal patient (19090). 
Concerning the short-term range it is possible to confirm a similar and 
perhaps normal deviation above unity that was also found in the top of Figure \ref{fig4_5}. In 
contrast, an unexpected feature is detected in the long-term range where it is 
possible to appreciate a deviation from a uniform power law. By analogy with 
the pattern of the simulation of $1/f$ fluctuations with the superimposed 
oscillatory trends (Figure \ref{fig2}), it is possible to suggest therefore for this 
patient the appearance of a very low frequency characteristic scale 
(reflected in the existence of a small but well defined projection in the 
long-term range of log 3 to 3.5). A larger than unity scaling exponent in the 
conventional DFA would approximately detect this deviation only if an 
appropriately reduced range (\textit{e.g.}, log $n$ from 3 to 3.5) for the linear slope 
estimation were selected. However, it would be impossible to select this 
corresponding range without further analysis and it is also clear that again a 
linear model, which revealed a short-term exponent of 1.12 ($n$ from 4 to 16) 
and a long-term exponent of 1.02 ($n$ from 100 to 10000), is not the best 
approximation for quantifying this type of scaling pattern. Hence, the 
analysis of the scaling pattern provided by the application of the $\alpha\beta$ filter confers the advantage, with no need of initial assumptions, of revealing for 
this case a clear behaviour that can be associated with a slow and weak 
oscillatory trend, or characteristic scale, in the HRV.

\begin{figure*}
\includegraphics{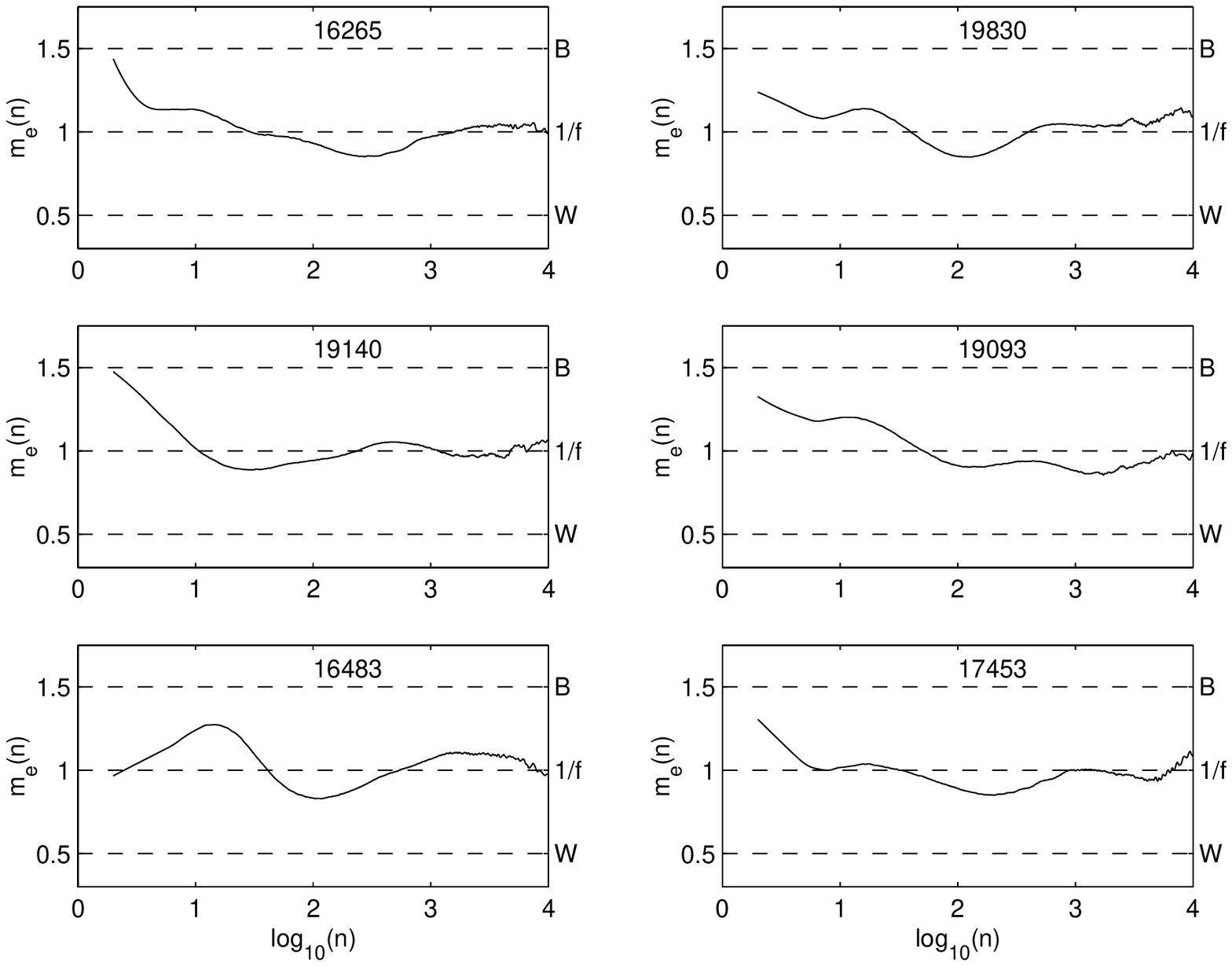}
\caption{\label{figcc}Scaling patterns $m_{e}(n)$ \textit{vs} log $n$ of normal patients apparently showing deviations above unity in the short-term range and uniform power law in the long-term range.}
\end{figure*}

\begin{figure*}
\includegraphics{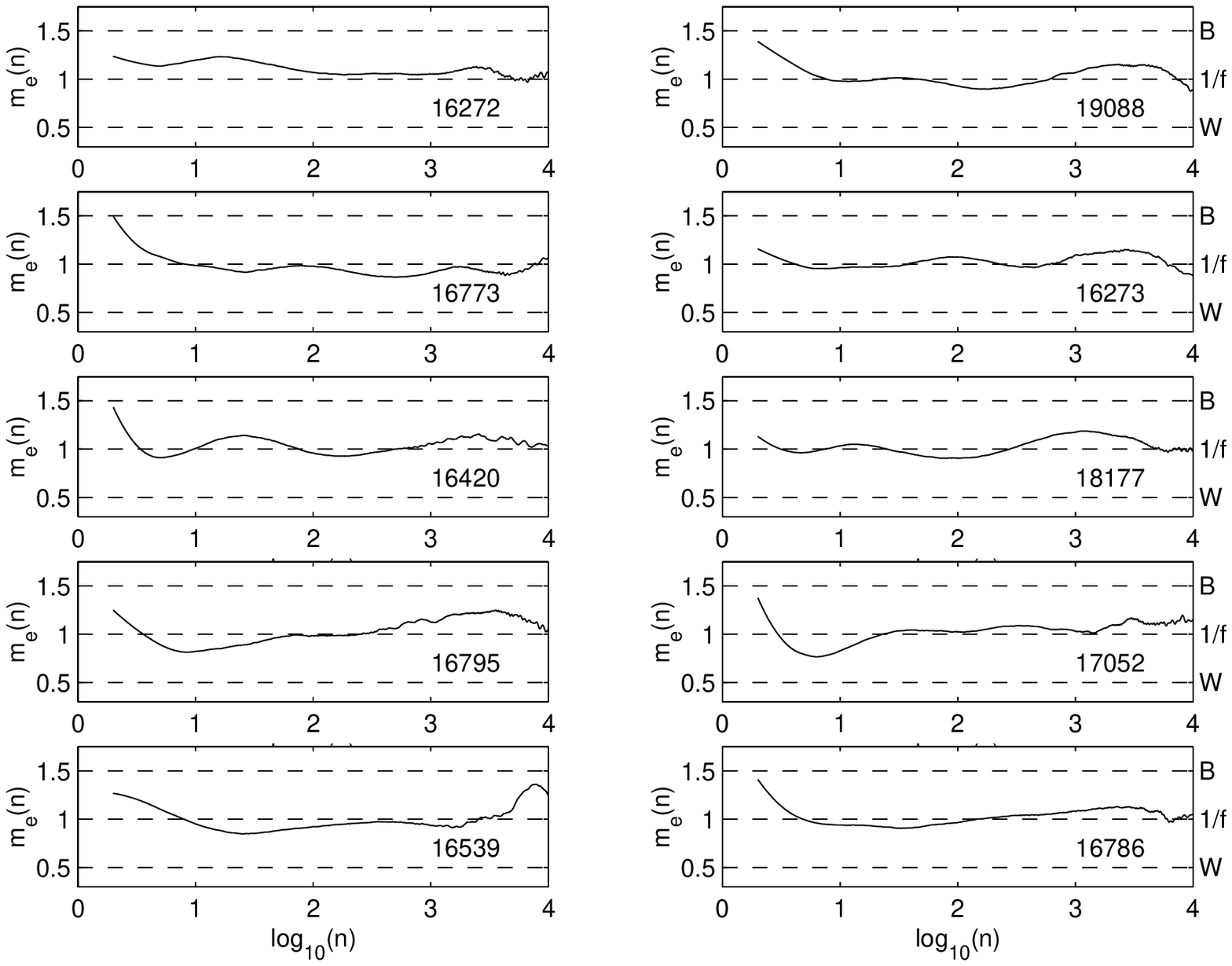}
\caption{\label{figdd}Scaling patterns $m_{e}(n)$ \textit{vs} log $n$ of normal patients not necessarily showing uniform power law in the long-term range or deviations above unity in the short-term range.} 
\end{figure*}

Concerning the rest of the normal group, Figure \ref{figcc} includes scaling patterns of another 6 patients apparently showing, like the top of Figure \ref{fig4_5}, both deviations above unity in short-term range and uniform power law in the long-term range. However, Figure \ref{figdd} presents the scaling patterns of the remaining 10 normal patients where it is also possible to appreciate, as at the bottom of Figure \ref{fig4_5}, that a uniform power law in the long-term range is not always manifested. Moreover, there are 3 cases in this figure (16420, 16795 and 17052) not necessarily showing the type of deviations above unity in short-term range similar to the ones depicted at the top of Figure \ref{fig4_5} or in Figure \ref{figcc}.

\begin{figure*}
\includegraphics{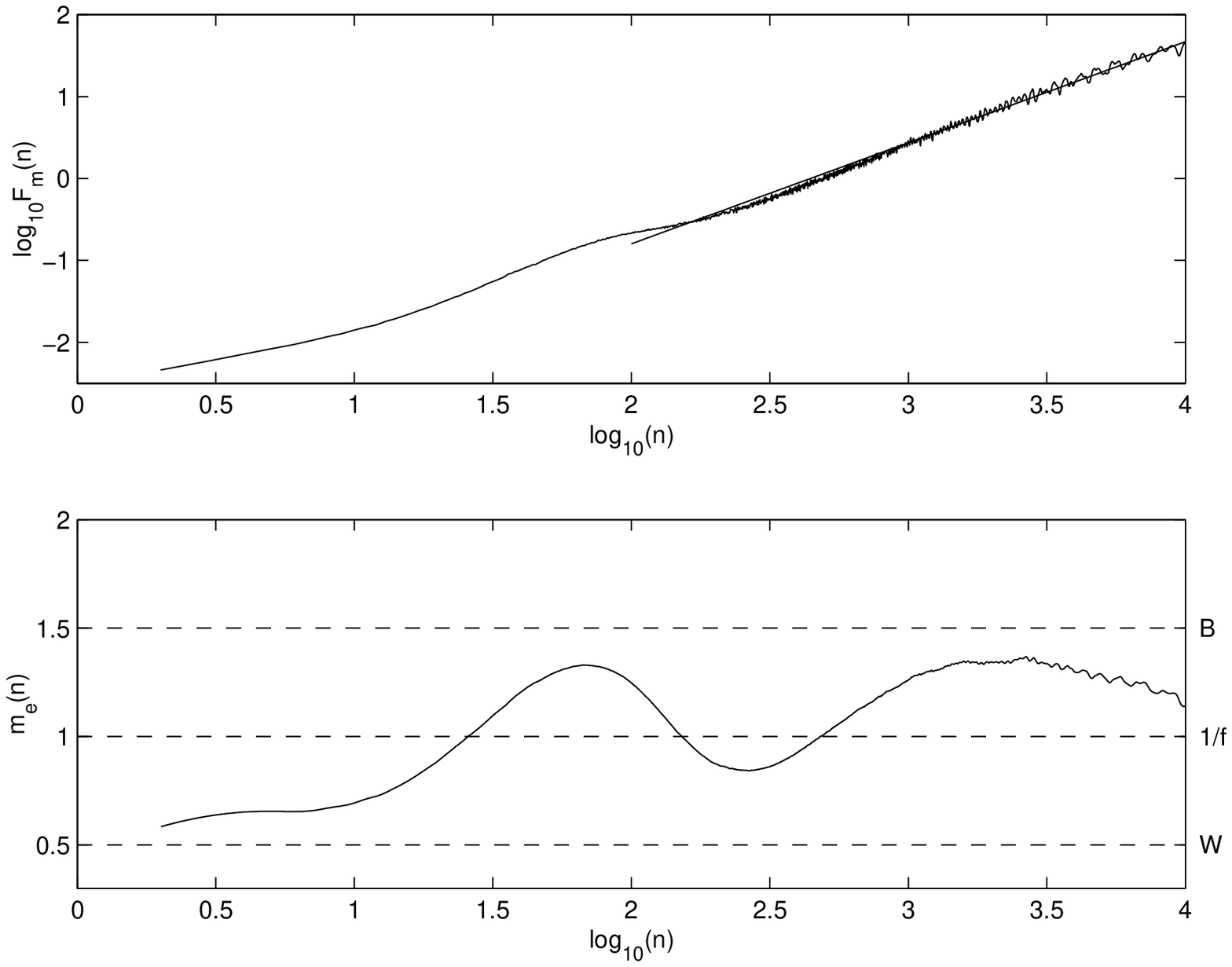}
\caption{\label{fig6}Result of DFA and $\alpha\beta$ filter applied to the RR interval series of Physionet CHF patient CHF03. The top plot shows log $F_{m}(n)$ \textit{vs} log \textit{n} obtained via DFA including a least-squares fit for 100$\le\quad n \quad\le$10000. The bottom plot 
presents $m_{e}(n)$ \textit{vs} log $n$ estimated by $\alpha\beta$ filter.}
\end{figure*}

Figure \ref{fig6} presents a typical scaling pattern for a CHF patient (CHF03). The 
pattern obtained by the $\alpha\beta$ filter clearly reveals 
major deviations, or projections, from a uniform power law in the intermediate and long-term range. In 
addition, a scaling pattern below unity, close to 0.5 (\textit{i.e.}, white noise), can be 
appreciated in the short-term range. The conventional DFA method has already 
revealed a very different behaviour in the scaling exponent of CHF patients 
\cite{4}; there, it was reported that for short time scales, the scaling 
exponents were below unity whilst for long-term range it was reported that the 
scaling exponents were above unity. The resulting scaling pattern of Figure \ref{fig6} 
provided by the application of the $\alpha\beta$ filter enhances the understanding of 
this abnormal behaviour. By analogy with the simulated results presented in 
Figures \ref{fig2} and \ref{fig3}, it is clear that the elevated value for the scaling exponent 
in the long-term range that was reported in Ref. \cite{4} is here shown in Figure \ref{fig6} 
to be related with the existence of predominant characteristic scales. 
Although the existence of a characteristic scale for the explanation of the 
abnormal scaling exponent was initially postulated in Ref. \cite{4}, the scaling 
patterns obtained by the $\alpha\beta$ filter confirm and complement this explanation. 
The major deviations of Figure \ref{fig6} seem to reveal not only one but two characteristic 
scales with potential physiological information that clearly cannot be 
properly detected by linear slope estimations over predefined ranges. This 
is illustrated in the top plot of Figure \ref{fig6} presenting the linear best fit 
for the range $n$ of 100 to 10000 that clearly misses the underlying pattern; 
resulting slope=1.23 . It seems necessary therefore to postulate that there 
could be more than one crossover phenomenon for the behaviour of this scaling pattern.

\begin{figure*}
\includegraphics{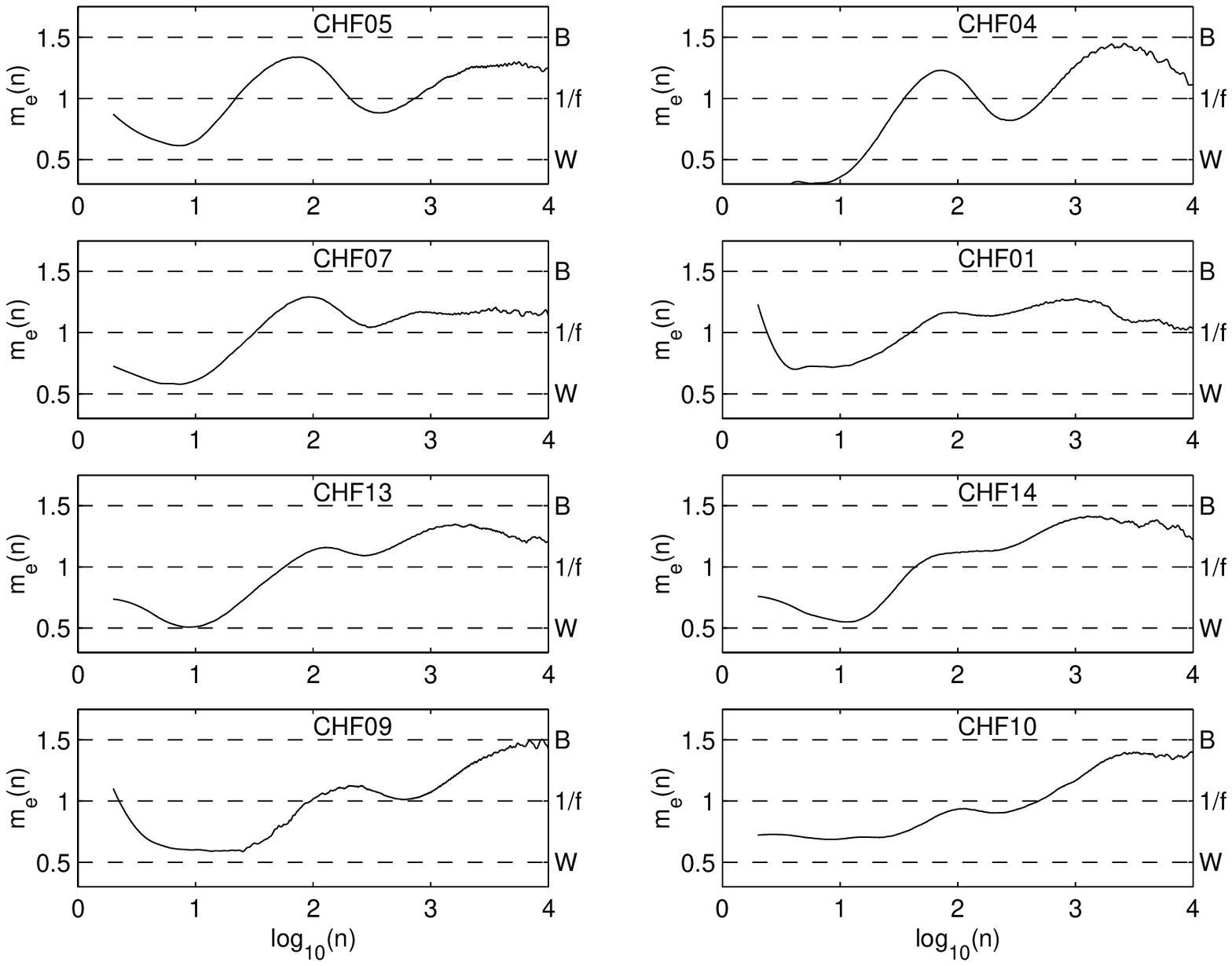}
\caption{\label{figaa}Scaling patterns $m_{e}(n)$ \textit{vs} log $n$ of CHF patients showing deviations below unity in the short-term range as well as characteristic scales in the intermediate and long-term range.}
\end{figure*}

It is important to report that for 8 additional CHF patients (Figure \ref{figaa}), the scaling patterns also presented deviations from a uniform power law that, as in Figure \ref{fig6}, suggest the existence of two characteristic scales in the intermediate and long-term ranges. These findings confirm the inadequacy of the conventional linear best fit to completely represent the underlying patterns revealed by DFA. In addition, like in Figure \ref{fig6}, it was also possible to observe in these 8 cases significant deviations below unity (toward 0.5) in the short-term range. This type of behaviour, similar to white noise, could be related to a marked attenuation of rhythmic oscillations \cite{22} probably indicating that in CHF the integrity of the autonomic mechanisms that regulate the RR intervals over short-time scales is affected. Additionally, this attenuation may also be related to poor end-organ responses \cite{Persson}.    

\begin{figure*}
\includegraphics{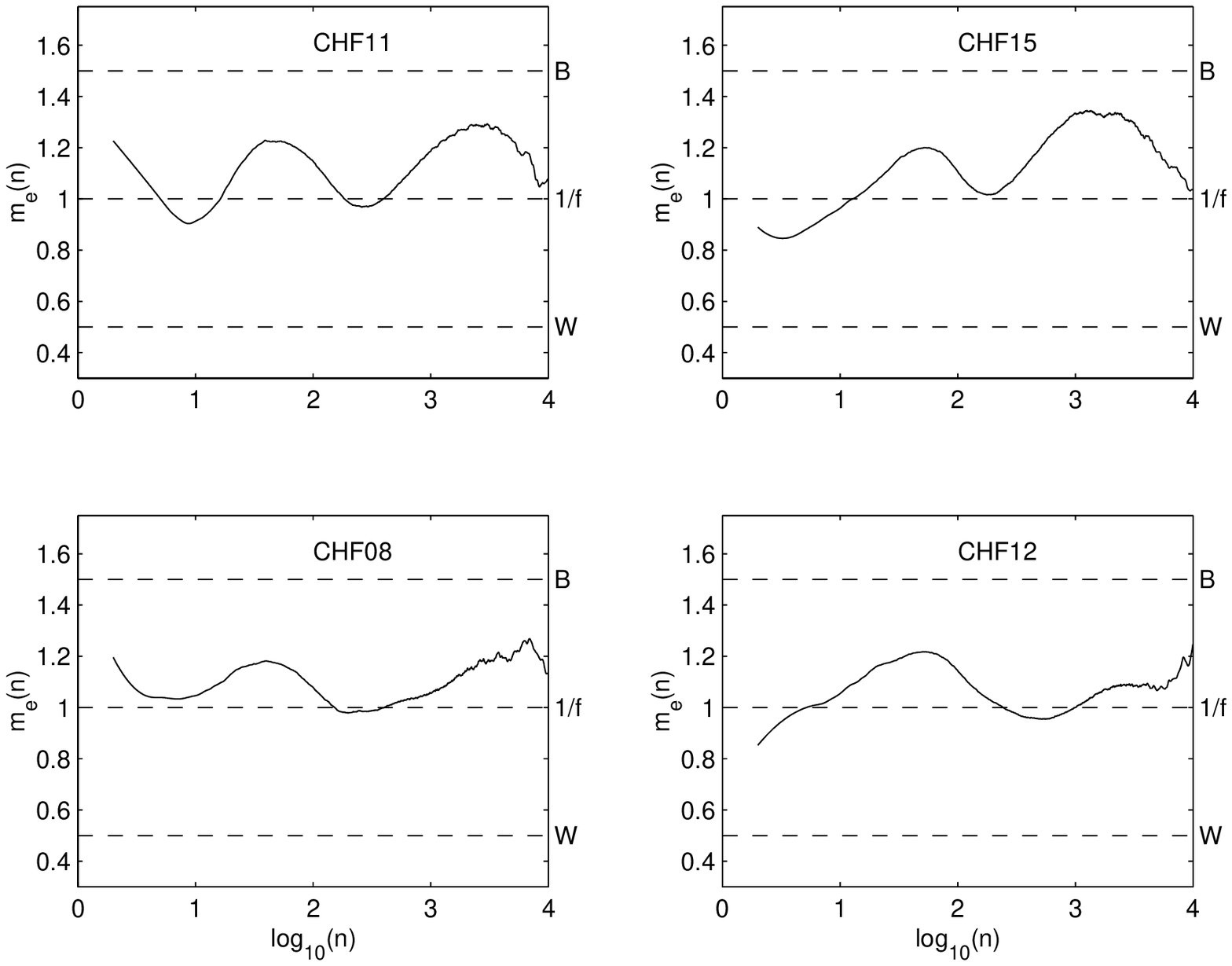}
\caption{\label{figbb}Scaling patterns $m_{e}(n)$ \textit{vs} log $n$ of CHF patients showing characteristic scales in intermediate and long-term range but not significant deviations below unity in the short-term range.}
\end{figure*}

Interestingly, the remaining CHF patients (Figure \ref{figbb}) also appear to present characteristic scales in the intermediate and long-term ranges but not significant deviations below unity in the short-term range. Hence, it could be speculated that the manifestation of CHF in these 4 patients was not as severe as in the cases presented in Figure \ref{figaa}. Consequently, this observation suggests that by performing a more detailed analysis of the scaling behaviour, as opposed to a conventional linear characterisation, it would be possible to recover information about the degree of severity of the CHF condition.

Having presented the results of CHF patients, it is then possible to describe additional features that were identified in the result of normal patients. In particular, scaling patterns 16420, 18177, 16795, and 17052 (Figure \ref{figdd}) appear to weakly show the type of characteristic deviations presented by the CHF group (figures \ref{figaa} and \ref{figbb}). Hence, it is speculated that the exploration of scaling patterns could also indicate the beginning of abnormalities, these perhaps expressed as subclinical manifestations of the cardio-respiratory control.

\begin{figure*}
\includegraphics{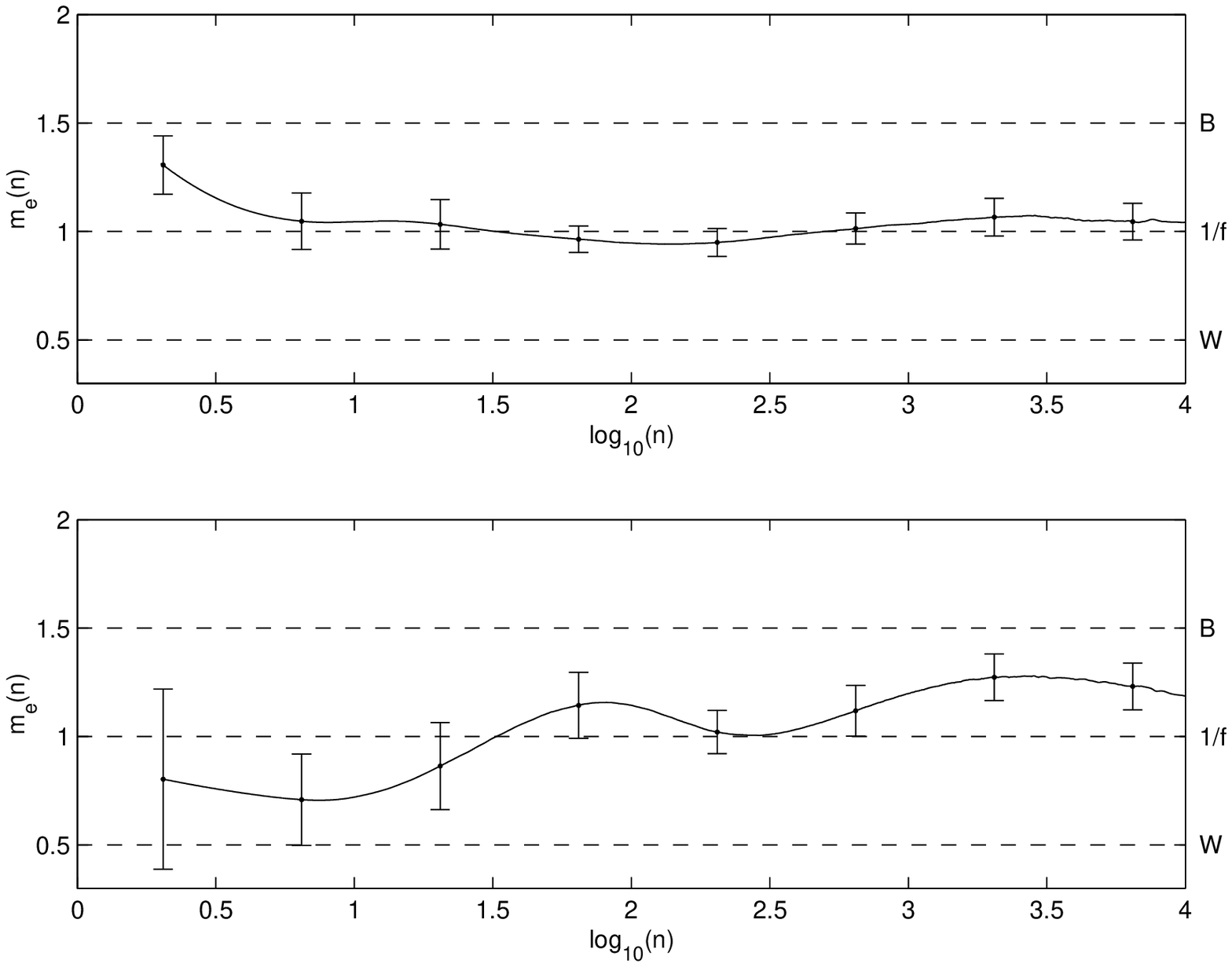}
\caption{\label{figee} Mean $\pm$ 1 standard deviation of the scaling patterns $m_{e}(n)$ \textit{vs} log $n$ obtained from the 18 cases in the normal group (top plot) and the 13 cases in the CHF group (bottom plot).}
\end{figure*}

To summarise, Figure \ref{figee} presents the mean scaling pattern obtained from the 18 cases in the normal group (Figures \ref{fig4_5}, \ref{figcc} and \ref{figdd}) as well as the same mean for the 13 cases in the CHF group (Figures \ref{fig6}, \ref{figaa} and \ref{figbb}). Also depicted in this figure are error bars (mean $\pm$ 1 standard deviation) at different log $n$ scales. Here, it is possible to appreciate that these scaling patterns have recovered the main features that we described for the normal and CHF groups (\textit{i.e.}, deviations above unity in the short-term range and uniform power law in the long-term range for normal patients, whereas CHF patients present deviations below unity in short-term range and two characteristic scales in the intermediate and long-term range). Interestingly, in spite of identifying in Figure \ref{figee} some scales showing an overlap in the error bars for both groups, it is also possible to detect particular scales without overlap in the short-term, intermediate, and long-term ranges so proving distinctive behaviour between groups. Additionally, the apparently larger error bars in the CHF scaling pattern may indicate the type of scaling instability in these patients that was first identified by Viswanathan \textit{et. al.}\cite{11}. 

\begin{figure*}
\includegraphics{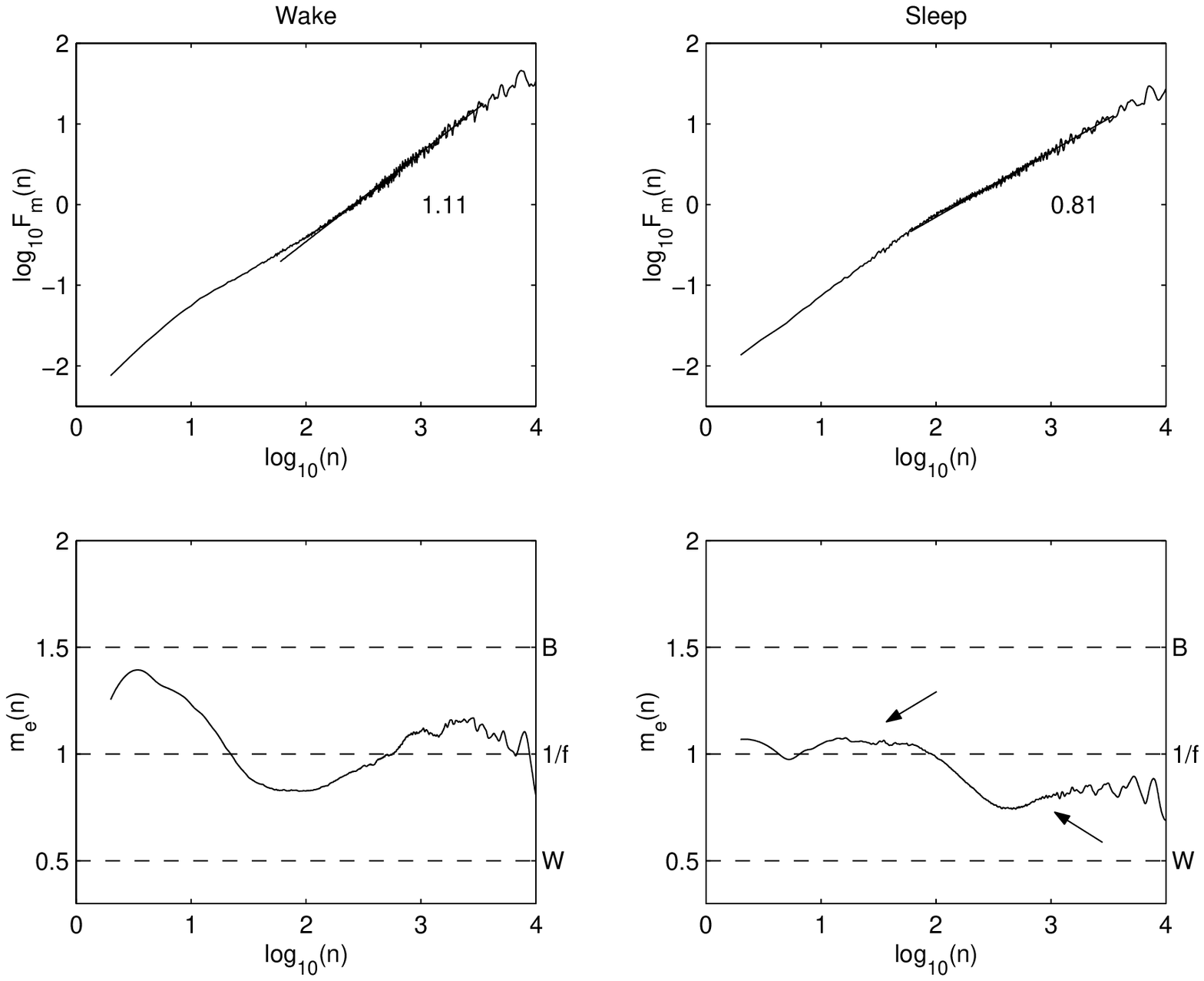}
\caption{\label{figff} Result of DFA and $\alpha\beta$ filter applied separately to wake (12-6 p.m.) and sleep (12-6 a.m.) periods of normal patient 16265. Top plots show log $F_{m}(n)$ \textit{vs} log \textit{n} obtained via DFA including a least-squares fit for 60$\le\quad n \quad\le$3500 as reported in Ref. \protect\cite{8}. Also indicated are the resulting scaling exponents. Bottom plots 
present corresponding $m_{e}(n)$ \textit{vs} log $n$ estimated by $\alpha\beta$ filter. Arrows indicate plateau in the intermediate range and different behaviour in long-term range.}
\end{figure*}

\begin{figure*}
\includegraphics{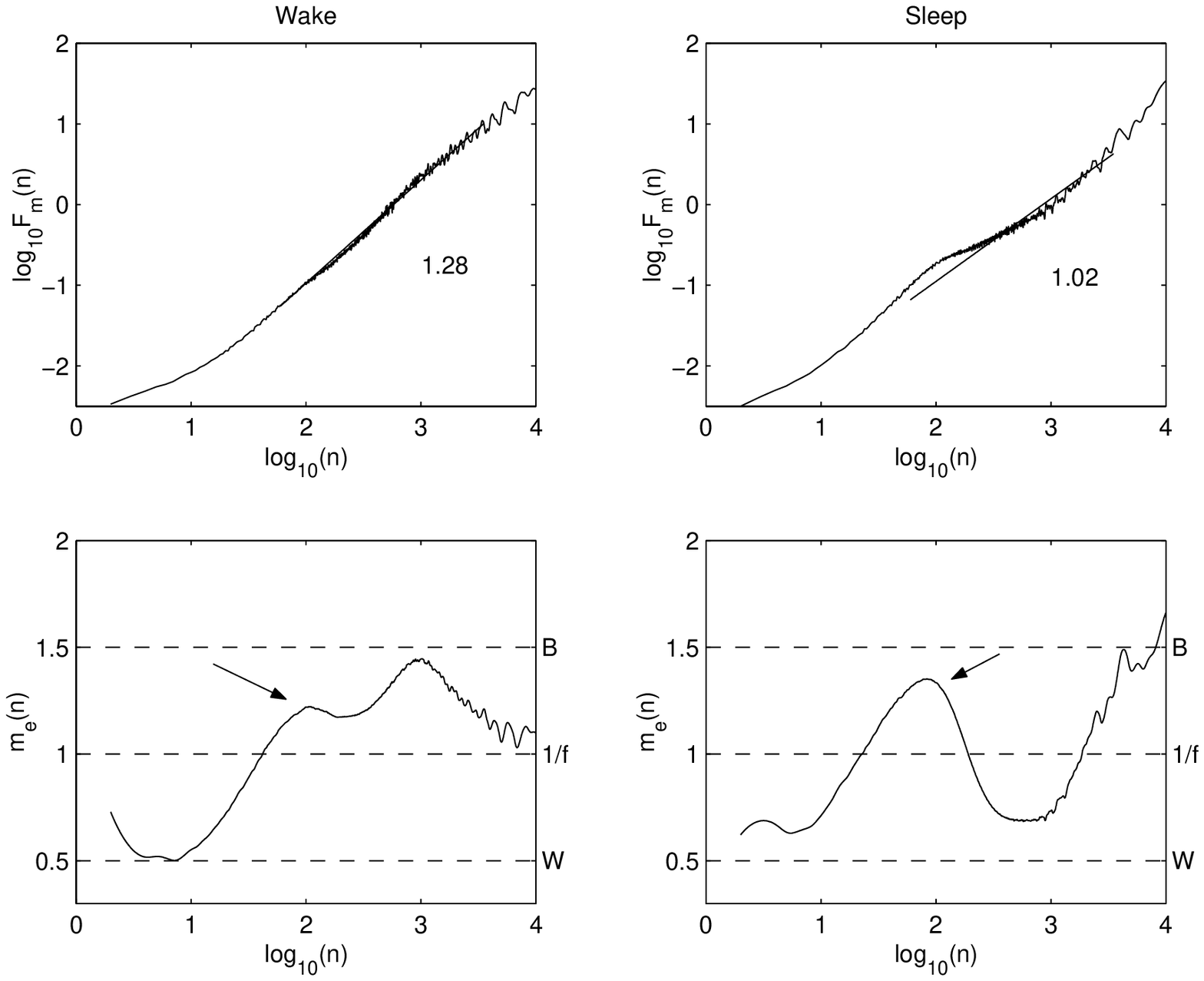}
\caption{\label{figgg} Result of DFA and $\alpha\beta$ filter applied separately to wake (1-7 p.m.) and sleep (12-6 a.m.) periods of CHF patient CHF07. Top plots show log $F_{m}(n)$ \textit{vs} log \textit{n} obtained via DFA including a least-squares fit for 60$\le\quad n \quad\le$3500 as reported in Ref. \protect\cite{8}. Also indicated are the resulting scaling exponents. Bottom plots 
present corresponding $m_{e}(n)$ \textit{vs} log $n$ estimated by $\alpha\beta$ filter. Arrows indicate the existence of intermediate characteristic scale.}
\end{figure*}

Finally, to show other examples of the advantages of analysing the scaling pattern without assuming predefined ranges or uniform 
power-law behaviour, figures \ref{figff} and \ref{figgg} present the results of the DFA and the $\alpha\beta$ filter applied separately to wake and sleep 6 hours segments of normal (16265) and CHF (CHF07) data. As reported in \cite{8}, using conventional quantification of the scaling behaviour (see top plots), it is possible to appreciate, both for the normal and CHF cases, sleep-wake differences in the long-term scaling exponents, finding smaller exponents during the sleep period. The scaling patterns at the bottom of the plots appear to provide additional information about these differences. 

For the normal case 16265 (Figure \ref{figff}) it is clear that there is a wake to sleep reduction in the scaling behaviour in the short-term range whereas for the intermediate range there is a marked scaling increase forming a type of plateau. Additionally, and this explains the reduction of the scaling exponent, there is a significant change of the long-term uniform power law toward a behaviour below unity. Clearly, the scaling pattern of this case obtained from the entire data set ($\simeq$ 24 hours) and shown in Figure \ref{figcc}, appears to include features of both wake and sleep segments. Hence, this pattern is not necessarily dominated by the scaling pattern of the wake segment, an observation that was suggested in Ref. \cite{8}, as a result of using a single exponent to characterise the data in the long-term range. On the contrary, this scaling pattern seems to be in accordance with results reported in \cite{Chen} showing that for signals comprised of segments with different local properties, the scaling behaviour is a superposition of the behaviour of different components. However, in that contribution \cite{Chen} it is also considered that the scaling behaviour is dominated by segments exhibiting higher positive correlation.

For the case CHF7 (Figure \ref{figgg}) it is possible to appreciate an intermediate characteristic scale in the scaling patterns of both sleep and wake segments. However, that projection appears to be more prominent during the sleep segment. Thus, this seems to be new information provided by the scaling patterns that cannot be inferred via the reduction of the scaling exponent (shown at the top of the plot). Clearly, the linear best fit misses the underlying pattern that reveals the existence of this characteristic scale around the log 2 range. Interestingly, according to this range and the mean RR interval, this characteristic scale having $\simeq$ 0.01 Hz could be the result of the abnormal heart rate oscillations associated with Cheyne-Stokes respiration that is commonly found in CHF patients \cite{4,Moody,Gold_Exp}. 

Concerning the long-term range of Figure \ref{figgg} it is possible to appreciate that the characteristic scale around the log 3 range which is present during wake period is no longer identified in the sleep segment. However, the scaling pattern of this segment shows a new deviation in the long-term range toward Brownian noise behaviour during the sleep segment. Once again, the scaling pattern of this case obtained from the entire record ($\simeq$ 20 hours) shown in Figure \ref{figaa} includes features of both wake and sleep segments.

The smaller values of the scaling exponent during the sleep segments shown in figures \ref{figff} and \ref{figgg} have been suggested in Ref. \cite{8} to be the result of a stronger neuroautonomic control during sleep periods. It remains to be seen in which sense this hypothesis is supported by the additional information described earlier from the scaling patterns. 

In this contribution several characteristic features have been detected in scaling patterns of HRV data of either normal or CHF patients. Hence, it is obviously difficult to suggest a single measurement to characterise these patterns. This circumstance may appear as a practical disadvantage of studying in detail the scaling behaviour using procedures like the $\alpha\beta$ filter. However, our results suggest that DFA involves potential information that appears to be ignored by not using a more suitable model to characterise the scaling behaviour. Consequently, this contribution may promote further studies involving larger groups aimed to obtain representative scaling patterns to typify different cohorts. Thus, new cases could be confronted with these representative patterns to determine the degree of similarity between them. Interestingly, this comparison could then be performed via single values obtained, for example, from measurements like the rms difference between candidates and representative patterns or from measurements obtained by a more sophisticated pattern recognition approach. These values would be useful to perform statistical comparisons or to assess severity of alterations in scaling patterns.           
                         
\section{Conclusion}

It has been suggested that the breakdown of persistent correlations over a wide range of time scales in HRV could be related to pathological conditions \cite{3}. The pattern analysis of the power-law behaviour that we present here via the incorporation of the $\alpha\beta$ filter supports and reveals strong deviations from a uniform power law in CHF conditions owing to a type of white noise behaviour and the existence of dominant characteristic scales. However, it is also suggested that caution should be taken about the existence of characteristic scales as precise indicators of nonhealthy dynamics. This is because some of the normal control patients also weakly presented characteristic scales in the long-term range, suggesting that either these types of deviations from a uniform power law are not necessary abnormal conditions or that these patients had subclinical manifestations that were not detected by the tests used to reveal cardiac abnormalities. 

Our results suggest the existence of variations in the DFA scaling pattern of the heart rate fluctuations with potential information that cannot be detected by linear-slope estimation procedures over predefined ranges. This exponent appears to be, consequently, a very rough approximation even for data from normal patients that are currently assumed to have in general a uniform power law in the long-term range. The incorporation of the $\alpha\beta$ filter appears as a convenient analytical improvement for the reliable estimation of these scaling patterns by avoiding initial assumptions or previous considerations about the nature of the data, or by avoiding a procedure like the one suggested by Hu \textit{et. al.} \cite{Hu} to find an optimal fitting range. It is argued in this work that this incorporation of the $\alpha\beta$ filter is a valuable tool for an appropriate analysis of complex data series like HRV, where the mechanisms underlying the power-law behaviour of these fluctuations still remain to be clearly elucidated. Hence, this technique will be of value in investigating whether, as some studies have suggested, endocrinal, hemodynamic, autonomic, and conscious states are factors involved \cite{10,23,24,25}, as well as the role played by the intrinsic nature of cardiac tissue \cite{26}.

\begin{acknowledgments}
The contribution of the Mexican Council for Science and Technology (CONACyT) for the financial support of J.C. Echeverr\'\i a is gratefully acknowledged.
\end{acknowledgments}

%\newpage

%\bibpreamble{\textbf{References}}

\bibliography{edfa_xarch}

\end{document}